\documentclass[10pt,conference]{IEEEtran} 
\usepackage[utf8]{inputenc}
\usepackage{graphicx}
\usepackage{amssymb}
\usepackage{amsmath}
\usepackage{amsfonts}
\usepackage{authblk}
\usepackage{amsthm}
\usepackage[english]{babel}
\usepackage{yhmath}
\usepackage{todonotes}
\usepackage{array}
\usepackage{csquotes}

\usepackage{url}
\usepackage[colorlinks=True, urlcolor=black, linkcolor=black, citecolor=black]{hyperref}

\usepackage[
backend=biber,
sorting=none,
style=numeric-comp,
bibstyle=numeric
]{biblatex}

\title{A Model for Circuit Execution Runtime And Its Implications for Quantum Kernels At Practical Data Set Sizes}

\author[1]{Travis L. Scholten\footnote{Corresponding author; Travis.Scholten@ibm.com}}
\author[2,**]{Derrick Perry II\footnote{This work began  while the author was an IBM Extreme Blue summer intern with the IBM Operations Risk Insights (ORI) group.}}
\author[3,**]{Joseph Washington}
\author[1]{Jennifer R. Glick}
\author[4]{Thomas Ward}
\affil[1]{IBM Quantum, IBM T.J. Watson Research Center, Yorktown Heights, NY 10598 USA}
\affil[2]{North Carolina Agricultural and Technical State University, Greensboro, NC 27411 USA }
\affil[3]{IBM Finance and Operations, Research Triangle Park, NC 27709 USA  }
\affil[4]{IBM Finance and Operations, Poughkeepsie, NY 12601 USA }
\affil[**]{Equal contribution}

\date{\today}

\begin{document}
\todototoc
\maketitle

\thispagestyle{plain}
\pagestyle{plain}
\begin{abstract}
    Quantum machine learning (QML) is a fast-growing discipline within quantum computing. One popular QML algorithm, quantum kernel estimation, uses quantum circuits to estimate a similarity measure (kernel) between two classical feature vectors. Given a set of such circuits, we give a heuristic, predictive model for the total circuit execution time required, based on a recently-introduced measure of the speed of quantum computers. In doing so, we also introduce the notion of an ``effective number of quantum volume layers of a circuit", which may be of independent interest. We validate the performance of this model using synthetic and real data by comparing the model's predictions to empirical runtime data collected from IBM Quantum computers through the use of the Qiskit Runtime service. At current speeds of today's quantum computers, our model predicts data sets consisting of on the order of hundreds of feature vectors can be processed in order a few hours. For a large-data workflow, our model's predictions for runtime imply further improvements in the speed of circuit execution -- as well as the algorithm itself -- are necessary.
\end{abstract}


\section{Introduction}

\begin{figure*}
\centering
\includegraphics[width=\linewidth]{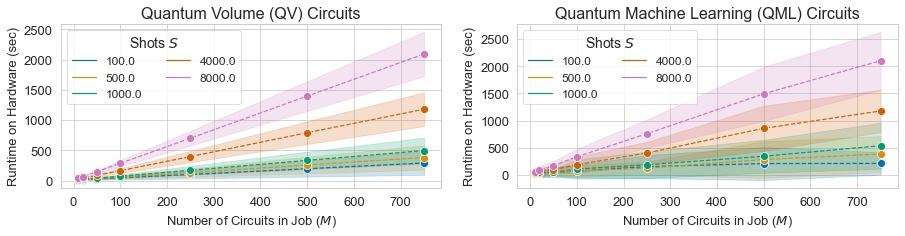}
\caption{\textbf{Empirical Runtimes of Circuits on Quantum Systems.} The plots above show the empirical runtime (in seconds) to execute a collection of $M$ quantum circuits (aka, a ``job"). The runtime generally increases with $M$, with a slope determined by the number of repetitions of each circuit (aka, ``shots", $S$), and other factors. In this work, we present a model for job runtime [Equation \eqref{eq:runtime-model}] based on a recently-introduced metric for measuring the speed of circuit execution, and evaluate its performance in predicting the runtime of quantum volume (QV) circuits and a particular circuit used in quantum machine learning (QML). Note that, for the quantum kernel estimation algorithm considered here, if a data set has $N$ elements, $M=\mathcal{O}(N^{2}).$ Shading indicates the standard deviation over the system used, and properties of the circuit itself (such as circuit width).}
\label{fig:empirical-runtimes}
\end{figure*}

Quantum machine learning (QML) is a broad, interdisciplinary topic at the intersection of quantum information/computation and classical machine learning \cite{dunjko2018machine, biamonte2017quantum,schuld2015introduction,cerezo2022challenges,ciliberto2018quantum,wittek2014quantum,dawid2022modern}. Within QML, there has been much study on one particular QML algorithm, called ``quantum kernel estimation" or ``quantum support vector machines" \cite{huang2021power,schuld2019quantum,havlivcek2019supervised,schuld2021supervised,gentinetta2022complexity,}. Quantum kernels are a similarity measure $K(\mathbf{x}, \mathbf{y})$ between two classical feature vectors (data points) $\mathbf{x}, \mathbf{y}$ evaluated using a quantum circuit\footnote{For details on kernel methods in general, see \cite{hofmann2008kernel, scholkopf2002learning}.}. This circuit uses an $n$-qubit parameterized encoding circuit $U(\boldsymbol{\theta})$. Given $U, \mathbf{x}$, and $\mathbf{y}$, and some fiducial starting state $|\psi_{0}\rangle$, the corresponding quantum kernel value is given by
\begin{equation}
\label{eq:q-kernels}
K(\mathbf{x}, \mathbf{y}) = |\langle \psi_{0}|U^{\dagger}(\mathbf{y})U(\mathbf{x})|\psi_{0}\rangle|^{2}.
\end{equation}
Usually, $|\psi_{0}\rangle$ is taken to be a computational basis state (typically, the all-zeros state, $|0^{\otimes n}\rangle$). To calculate a quantum kernel using a quantum computer, $|\psi_{0}\rangle$ is prepared, and the circuit $U(\mathbf{x})\circ U^{\dagger}(\mathbf{y})$ is applied. (Here, $\circ$ means the composition of two operators.) Finally, the resulting state is measured, resulting in a classical bitstring $\mathbf{b}$. The probability of obtaining the bitstring corresponding to $|\psi_{0}\rangle$ is estimated by repeating the just-described process many times (aka, for many ``shots") to build up statistics:
\begin{equation}\label{eq:q-kernels-estimate}
\widehat{\mathrm{Pr}(|\psi_{0}\rangle)} = \frac{\text{\# of outcomes $\mathbf{b}$ corresponding to $|\psi_{0}\rangle$}}{S},
\end{equation}
with $S$ as the number of shots. Here, the symbol ``~$\widehat{}$~" is used in the statistical sense of ``Is an estimate of", not in the quantum-mechanical sense of ``Is a quantum-mechanical operator". That is, Equation~\eqref{eq:q-kernels-estimate} is an estimate of the quantum kernel, Equation \eqref{eq:q-kernels}.

Given a data set $\mathcal{D}  = \{\mathbf{x}_{1}, \mathbf{x}_{2}, \cdots , \mathbf{x}_{N}\}$, usually the collection of pairwise quantum kernel values $K(\mathbf{x}_{1}, \mathbf{x}_{1}), K(\mathbf{x}_{1}, \mathbf{x}_{2}), \cdots $ is estimated. These values can then be used in  classical kernel-based algorithms, such as support vector machines \cite{cristianini2000introduction}, Gaussian processes \cite{williams2006gaussian},  etc. \cite{jacot2018neural,vishwanathan2010graph}. In this way, quantum kernels ``enhance" classical kernel-based algorithms. This work focuses on quantum-enhanced support vector machines.

Quantum kernels have already been used in a variety of contexts, including high-energy physics \cite{fadol2022application,belis2021higgs, wu2021application}, healthcare and life sciences \cite{mensa2022quantum,krunic2022quantum}, many-body physics \cite{sancho2022quantum}, natural language processing \cite{wright2022design}, industrial manufacturing \cite{beaulieu2022quantum}, and financial services and banking \cite{kyriienko2022unsupervised}. However, to date the only proof of an advantage from using quantum kernels is theoretical in nature \cite{liu2021rigorous}. In a practical context, quantum advantage with quantum kernels has yet to be attained.

One obstacle to deploying quantum kernels in practice -- and at scale across a data set where $N >> 1$ -- is the time spent executing the necessary quantum circuits could become a bottleneck to the total runtime of the quantum-enhanced, kernel-based algorithm. At least two places exist where this bottleneck could arise: first, transferring data to the quantum computer (necessary because, usually, quantum computers are not closely co-located with the data sets they are processing, necessitating the transfer of data over networks), and second, the total time required for the quantum computer to run the required circuits. The former obstacle can be alleviated by minimizing the amount of data transfer required \cite{naveh2021kernel}; the latter is the subject of this work. The question we consider here is: ``How much time is needed to execute a job consisting of $S$ shots each of $M$ circuits, each of which estimates a quantum kernel value based on an encoding circuit $U(\boldsymbol{\theta})$?".

The runtime must clearly relate to $M$ itself, as well as $S$, as evidenced by Figure \ref{fig:empirical-runtimes}. However, other properties of the circuit itself -- as well as the system the job is being run on -- may also impact runtime.  In this work, we introduce a well-motivated model for job runtime (Section \ref{sec:runtime-model}), and evaluate its performance by comparing the model's predictions to results obtained from running jobs on IBM Quantum's systems using the Qiskit Runtime \cite{QiskitRuntime} service (Section \ref{sec:model-performance}). Using this model, we then discuss the implications of estimating quantum kernels on practical and large data sets in for a climatologically-relevant problem; namely, flash flood prediction (Section \ref{sec:model-extrapolation}). Finally, we conclude with 
a discussion of the implications of the model for processing large data sets ($N>>1$), as well as interesting directions for future work (Section \ref{sec:conclusions}).


\section{A Model for Circuit Execution (Job) Runtime}
\label{sec:runtime-model}

This section presents a model for job runtime. It model does \emph{not} take into account the time a given job spends waiting in a queue prior to being executed on hardware. Empirical studies of queue times show wide variation in how long a given circuit spends waiting to execute; see \cite[Fig 3]{ravi2021quantum}. Queue time depends strongly on the queuing system used; instead, this work focuses on modeling the time required to run the job once it has been removed from the queue.

Modeling job runtime is hindered due to a lack of well-defined notions of ``How long does it take a quantum computer to run a circuit?". One starting point is using information about how much time is needed for state initialization, gates, and measurements. However, such a model may be overly-cumbersome to use in practice, as modeling the runtime of a circuit with even a moderate number of qubits or depth could be difficult. Doing so would require getting down into the weeds of the circuits, and considering the vagaries of how the hardware executes them\footnote{For example, whether the compilers used to schedule pulses attempt to bring pulses forwards in time in the pulse-based representation of the circuit.}. What's more, such a low-level model misses the impact of contributions higher up the stack on timing performance -- for example, the time spent compiling an abstract quantum circuit or program into the requisite pulse signals would clearly impact overall runtime, but wouldn't be captured by such a model.

Hence, a better model -- in the sense of capturing more of the stack that impacts timing performance -- would focus on modeling runtime starting from the moment a given job is pulled from a queue of jobs, to the time its results are sent back to the end-user. The necessary ingredient to do so is a \emph{holistic} notion of ``system speed".

Such a quantity has been recently introduced in the literature, and is called ``Circuit Layer Operations Per Second" (CLOPS) \cite{wack2021quality}. The methodology used to calculate the CLOPS of a given system explicitly encompasses the entire stack from the moment a job is de-queued, and is straightforward to describe. Consider running a job of $M$ parameterized quantum volume circuits \cite{cross-qv-2019} on a system with quantum volume $V$. Each circuit in the job has a number of quantum volume \emph{layers} (repetitions of permutations and random 2-qubit gates) $D=\log_{2}(V)$. And suppose the parameters of each circuit are updated updated $K$ times, and each circuit in the job is repeated for $S$ shots. Let the total elapsed time be $T$. The CLOPS $C$ of the system is then
\begin{equation}
\label{eq:CLOPS-def}
C = \frac{MDKS}{T}.
\end{equation}
The methodology for computing CLOPS presented takes $S=M=100$, $K=10$, and performs the parameter updates by chaining the output of one run of a circuit to the inputs of the next run, through the use of a pseudo-random number generator \cite{wack2021quality}.

Assuming the stack has no fixed overheads or time costs with respect to varying any of $M,K,S$, or $D$, then a multiplicative scaling of any of these parameters would result in a corresponding scaling of the total runtime. That is, if another job was run with $M'$ circuits, $K'$ parameter updates, $S'$ shots, and $D'$ quantum volume layers, then a system with CLOPS $C$ \textit{should} take a time
\begin{equation}
\label{eq:clops-scaled}
T' = (M'*K'*D'*S')/C
\end{equation}
to run such a job.

To apply Equation \eqref{eq:clops-scaled} to jobs consisting of circuits which estimate quantum kernel values, two modifications are necessary. Both relate to the fact the CLOPS metric is defined using quantum volume circuits, but  quantum volume circuits are not usually used as encoding circuits in QML.

The first -- and most straightforward -- issue is the CLOPS metric incorporates the notion of parameter updates through the variable $K$. When calculating quantum kernels, no parameter updates are done; $K$ should be fixed to one\footnote{Note if quantum kernel training \cite{glick2021covariant} was performed, then $K\neq 1$, and should reflect the number of update calls performed.}.

\begin{figure}
\centering
\includegraphics[width=.8\linewidth]{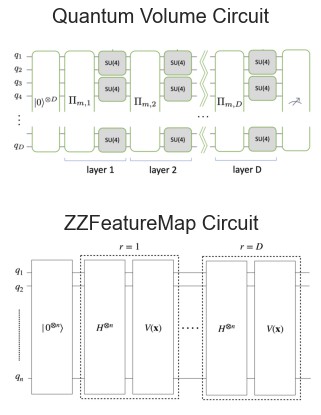}
\caption{\textbf{Notion of Number of Repetitions of a Base Template for a Circuit.} The circuits considered here use a number of repetitions $D$ of a \emph{base template}. \textbf{Top:} For quantum volume circuits, the base template consists of a permutation of the qubits, followed by random 2-qubit gates. \textbf{Bottom:} For quantum kernel circuits based on the ZZFeatureMap [Equation \eqref{eq:zzfm-circuit}], the base template consists of a layer of Hadamard gates, and the application of a parameterized circuit $V(\mathbf{x})$. NOTE: The image for the quantum volume circuit is taken from \cite[Fig 3]{wack2021quality}.}
\label{fig:base-templates}
\end{figure}

The second issue is what the notion of ``number of quantum volume layers" ($D$) would mean. While a given feature map may have a parameter which seems similar in spirit to $D$ -- for example, by repeating a \emph{base template} for an encoding circuit several times -- these are different categories of items, making them incomparable. Figure \ref{fig:base-templates} shows examples of what both ``number of repetitions of a base template" mean for quantum volume and a particular QML circuit, called a ``ZZFeatureMap" ([Equation \eqref{eq:zzfm-circuit}] and reference \cite{havlivcek2019supervised}).

Consequently, a notion of the ``effective" number of quantum volume layers is needed. We provide a definition below, based on 2 observations. The first  observation is for an $n$-qubit encoding circuit $U(\mathbf{x})$, with a number of repetitions of its template $D$, the corresponding circuit for calculating a quantum kernel acts on $n$ qubits and has a number of repetitions of the base template $2D$. Thus, its \emph{volumetric area}\footnote{The notion of volumetric area of a circuit is based on the idea of volumetric benchmarking of quantum computers \cite{blume2020volumetric}, with the difference that in \cite{blume2020volumetric}, the depth of the circuit when transpiled to a canonical gate set is used in place of a notion of ``number of layers".} -- the product of circuit width and number of base layers -- is $2Dn$. A quantum volume circuit acting on $q$ qubits has volumetric area\footnote{Recall quantum volume circuits are \emph{square}, meaning the number of quantum volume layers is equal to the number of qubits the circuit acts on.} $q^{2}$. Thus, a quantum volume circuit with $q^{2} = 2Dn$ has the same volumetric area as a quantum kernel circuit. This sets a required number of qubits the quantum volume circuit needs to act on in order to have the same volumetric area as the quantum kernel circuit.

\begin{figure}
\centering
\includegraphics[width=.9\linewidth]{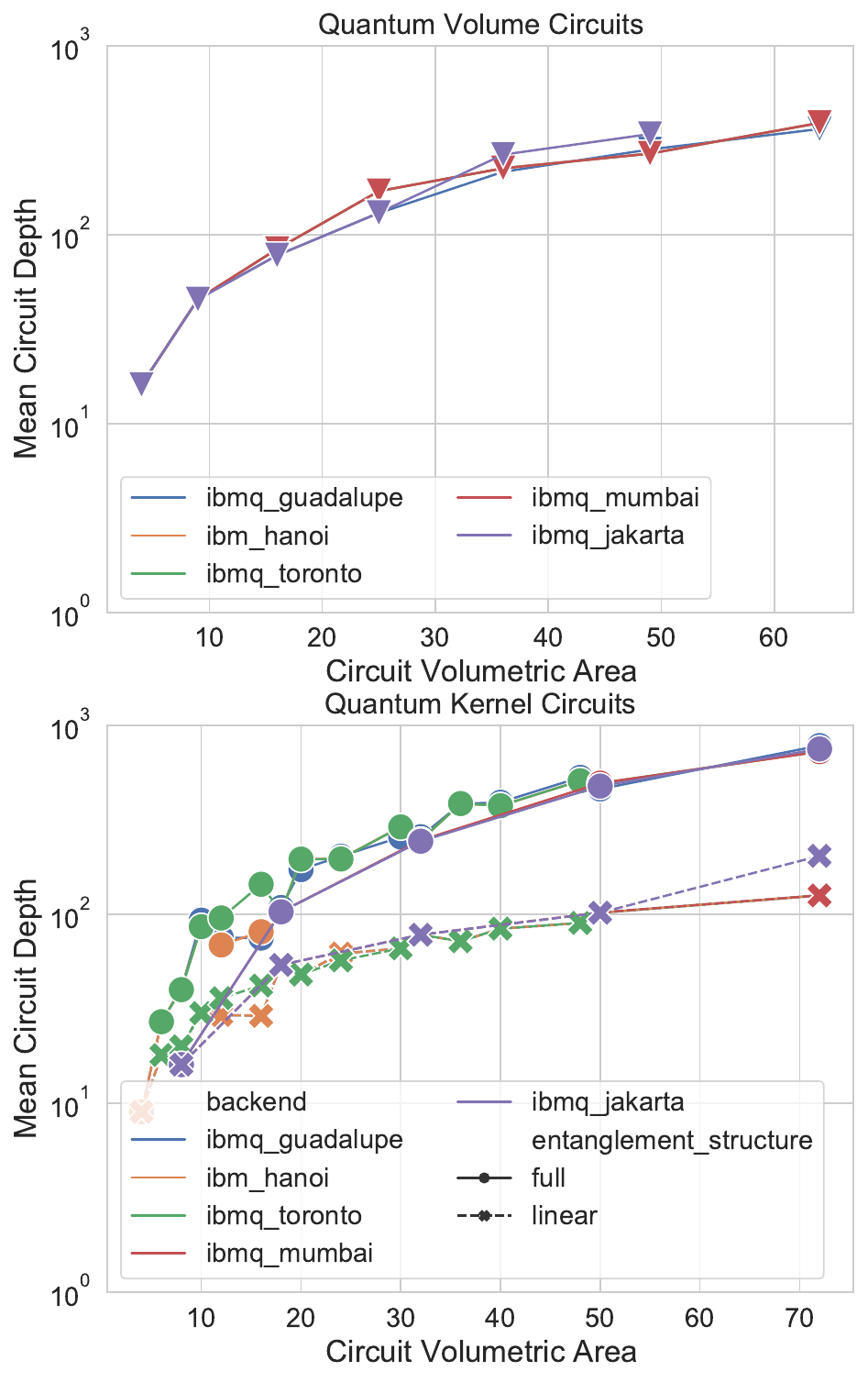}
\caption{\textbf{Circuits with the Same Volumetric Area Do not Have the Same Depth.} Comparing the volumetric area (product of circuit depth and number of repetitions of a base layer) for quantum volume (QV) circuits (\textbf{top}) and quantum kernel circuits based on the ZZFeatureMap (\textbf{bottom}; Equation \eqref{eq:zzfm-circuit}) shows that when transpiled to hardware, circuits with the same volumetric area can have dramatically different depths. The entanglement structure -- arrangement of two-qubit gates -- for the ZZFeatureMap circuits dramatically impacts depth. Differences in depth will translate into differences in circuit execution time, and motivate the use of a depth-dependent prefactor in Equation \eqref{eq:eff-d}. Note: For QV circuits, the mean is calculated over 20 random realizations, and for the ZZFeatureMap circuits, at least 25 random realizations are used. }
\label{fig:vol-area-depths}
\end{figure}

The second observation is even when two circuits have the same volumetric area, their depths when transpiled to hardware will generally not be the same (see Figure \ref{fig:vol-area-depths}). A variety of circuits with different values of $n$ and $D$ can have the same volumetric area, but the circuit execution time can be dramatically different -- intuitively, a circuit with higher depth will take more time to execute. Hence, capturing the effect of circuit depth is necessary. To do so, we normalize the depth of the quantum kernel circuit to the depth of a quantum volume circuit with the same volumetric area, and use it as a scaling factor.

These two observations above lead to a definition of the ``effective number of quantum volume layers" of a quantum kernel circuit as
\begin{equation}
\label{eq:eff-d}
D_{\mathrm{eff}} \equiv \frac{\langle \mathrm{Depth}(U^{\dagger}(\mathbf{x})U(\mathbf{y}))\rangle}{\langle \mathrm{Depth}(QV_{v}) \rangle}*v,~\text{where}~v =  \left \lceil \sqrt{2Dn} \right \rceil.
\end{equation}
Here, $QV_{j}$ denotes a quantum volume circuit with a number of layers $j$, and $\mathrm{Depth}()$ denotes the circuit depth when transpiled onto hardware. The expectation values are taken with respect to the parameters $\mathbf{x}, \mathbf{y}$ and random seeds for the kernel and quantum volume circuits, respectively.

Thus, our model for execution time of a job consisting of $M$ quantum kernel circuits with an effective number of quantum volume layers $D_{\mathrm{eff}}$ on a system with with CLOPS $C$ for a total of $S$ shots is given by
\begin{equation}
\label{eq:runtime-model}
\hat{T} = \frac{MS}{C}*D_{\mathrm{eff}}.
\end{equation}
Note here, $\hat{T}$ means ``An estimate of the runtime", not ``Is a quantum-mechanical operator".


\section{Model Performance}
\label{sec:model-performance}

The performance of the model is evaluated using 2 kinds of circuits: quantum volume circuits and kernel circuits based on the ZZFeatureMap circuit. Both of these circuits are parameterized, so synthetic data is generate the parameters. Empirical runtime information is collected by submitting the jobs to IBM Quantum systems using the Qiskit Runtime, a quantum computing service and programming model allowing users to optimize workloads and efficiently execute them on quantum systems at scale \cite{QiskitRuntime}, via the Runtime's Sampler primitive \cite{QiskitSampler}. Across the jobs, the number of circuits $M$, shots $S$, backend used, and number of qubits $n$ are varied. In addition, for the ZZFeatureMap circuits, both the number of repetitions of the base template $D$ and the circuit's entanglement structure are varied.

To quantify the model's performance at predicting runtime, two numbers are used. Suppose the actual runtime for the job is $T$, and the runtime predicted by the model is $\hat{T}$. The corresponding \emph{loss} $L$ of the model with respect to the job is be
\begin{equation}
\label{eq:loss}
L = \begin{cases}r-1~~~~~~r \geq 1 \\ 1/r-1~~~r<1\end{cases}~~\text{with}~~r = \frac{\hat{T}}{T},
\end{equation}
By construction $L \geq 0$, with equality if, and only if, the predicted and actual runtimes agree. 
The number $r$ -- the \emph{runtime ratio} -- is another quantifier of the degree to which the predicted and actual runtime agree. When $r < 1$, the model \emph{under-predicts} runtime. One problem with this is if the predictions of the model are used in other contexts -- for example, analyzing the overall runtime of a QML workflow -- then an under-prediction on the part of the model would negatively impact such an overall analysis. Hence, the loss function more strongly penalizes under-prediction of runtime (i.e., increases more quickly  when $r < 1$).

\subsection{Model Performance: Quantum Volume Circuits}

\begin{table}

\centering
\caption{Model Performance for Quantum Volume circuits (CLOPS Job)}
\label{tab:clops-exp-design-qv}

\begin{tabular}{|m{7em}|m{5em}|m{6.5em}|m{3.5em}|m{2em}|}
\hline
      \textbf{ Backend (QV, CLOPS)} &  \textbf{Actual runtime $T$ (s)} &  \textbf{Predicted Runtime $\hat{T}$ (s)}&  \textbf{Runtime Ratio $r$} &  \textbf{Loss $L$ }\\ \hline
     ibm\_hanoi (64, 2.3K) &            68.0 &               25.6 &            0.4 &   1.7 \\ \hline
ibmq\_guadalupe (32, 2.4K) &            41.0 &               21.3 &            0.5 &   0.9 \\ \hline
  ibmq\_jakarta (16, 2.4K) &            31.0 &               16.4 &            0.5 &   0.9 \\ \hline
   ibmq\_mumbai (128, 1.8K)  &            97.0 &               38.2 &            0.4 &   1.5 \\ \hline
  ibmq\_toronto (32, 1.8K) &            48.0 &               28.0 &            0.6 &   0.7 \\
\hline
\end{tabular}

\end{table}

The runtime model uses the CLOPS metric as the notion of the speed of circuit execution. The CLOPS metric is computed using quantum volume circuits. Hence, we first evaluate the performanc of the model when the circuits in the job are quantum volume circuits. Note for these jobs, $D_{\mathrm{eff}}$ in Equation \eqref{eq:runtime-model} is taken to be the number of quantum volume layers $D$.

Table \ref{tab:clops-exp-design-qv} shows --  across 5 backends -- the actual runtime $T$, the runtime predicted by the model $\hat{T}$, the runtime ratio $r=\hat{T}/T$, and the corresponding loss for jobs where $S=M=100$, and $D=$ log$_2$(QV). (Note  in these experiments $K=1$, whereas for the CLOPS experiments, $K=10$.) The value of the runtime ratio $\hat{T}/T$ shows the model consistently \emph{under-estimates} the runtime. As a result, the model's loss is non-zero. One reason for this discrepancy could be that, when calculating a system's CLOPS, the quantum volume circuits are \emph{pre-transpiled} to a given system. For the jobs submitted here, they \emph{were not}, meaning some additional time was spent in transpilation.

Note the number of quantum volume layers $D$ depends on the quantum volume of the backend; the circuits run on systems with higher quantum volumes have more layers than those run on systems with lower quantum volumes. Hence, even if 2 systems have roughly the same CLOPS values, their actual runtimes may be different, due to differences in the number of layers in the circuit. Further, different systems have different numbers of qubits, which impacts the time cost of circuit transpilation and waveform loading. Thus, even though \emph{ibmq\_jakarta}, \emph{ibmq\_guadalupe}, and \emph{ibm\_hanoi} all have a comparable CLOPS value, their differences in quantum volume and qubit count mean the actual runtime $T$ will be different for these CLOPS jobs.

\begin{table*}

\centering

\caption{Model Performance for Quantum Volume Circuits (CLOPS-like job).}
\label{tab:clops-exp-design-qv2}


\begin{tabular}{|l|c|c|c|c|c|c|c||c|c|c|c|c|c|c|}
\hline
{} & \multicolumn{7}{l}{\textbf{Loss} $L$} & \multicolumn{7}{l}{\textbf{Runtime Ratio} $r$}  \\ \hline
\textit{Shots, S} &   10   &  50   &  100  &  500  &  1000 &  4000 &  8000 &          10   &  50   &  100  &  500  &  1000 &  4000 &  8000 \\
\textbf{Backend}        &        &       &       &       &       &       &       &               &       &       &       &       &       &       \\ \hline
ibm\_hanoi      &  24.36 &  4.07 &  1.65 &  \textbf{0.60} &  1.76 &  4.51 &  5.72 &          0.04 &  0.20 &  0.38 &  1.60 &  2.76 &  5.51 &  6.72 \\ \hline
ibmq\_guadalupe &  16.87 &  2.57 &  \textbf{0.93} &  1.01 &  2.08 &  4.22 &  4.97 &          0.06 &  0.28 &  0.52 &  2.01 &  3.08 &  5.22 &  5.97 \\ \hline
ibmq\_jakarta   &  16.68 &  2.78 &  \textbf{0.89} &  0.91 &  1.83 &  3.46 &  3.95 &          0.06 &  0.26 &  0.53 &  1.91 &  2.83 &  4.46 &  4.95 \\ \hline
ibmq\_mumbai    &  23.34 &  3.71 &  1.54 &  \textbf{0.72} &  1.99 &  5.59 &  7.20 &          0.04 &  0.21 &  0.39 &  1.72 &  2.99 &  6.59 &  8.20 \\ \hline
ibmq\_toronto   &  16.49 &  2.43 &  \textbf{0.71} &  1.26 &  2.59 &  5.67 &  6.84 &          0.06 &  0.29 &  0.58 &  2.26 &  3.59 &  6.67 &  7.84 \\ \hline
\end{tabular}

\end{table*}

The methodology used to compute CLOPS uses $S=100$. This is a small value for applications where precise estimates are required; commonly, jobs use on the order of thousands of shots. For quantum kernels, increasing the number of shots directly increases the accuracy with which the kernel [Equation \eqref{eq:q-kernels}] can be estimated. And, as shown back in Figure \ref{fig:empirical-runtimes}, changing $S$ dramatically changes the runtime.

This is also reflected in the results of Table \ref{tab:clops-exp-design-qv} which extend Table \ref{tab:clops-exp-design-qv2} to run the exact same set of jobs, except the number of shots is changed. Considering the model's loss for these jobs, we see it is minimized when $S$ is 100 or 500 -- exactly (or close to) the number of shots used for measuring CLOPS.

As $S\rightarrow 0$ the loss increases substantially, because the runtime ratio approaches $0$, driven by the fact that in the model [Equation \eqref{eq:runtime-model}] the number of shots enters multiplicatively in the predicted runtime. However, there are fixed overheads across the stack which don't scale with $S$. For example, as noted in \cite{wack2021quality}, the time required for circuit compilation and data transfer is independent of $S$. Such an overhead would dominate circuit runtime in a low-shot regime.

When the number of shots increases, the loss does as well, albeit less dramatically as when the number of shots decreases. In terms of the runtime ratio, as $S$ increases, the model \emph{over-predicts} job runtime, though the runtime ratio appears to be similar across similar backends.

These results imply that although the model is not perfectly accurate with respect to predicting runtimes for the CLOPS job, it is -- comparatively speaking -- most accurate for such a (or a very similar) job, as opposed to jobs involving a small or large number of shots. As we will discuss in Section \ref{sec:conclusions}, one of the main reasons for these discrepancies could be the fact  the CLOPS metric is evaluated using an execution path different from the one used here. That is, the manner in which jobs are set up and run is different, which can lead to differences in execution time, a point returned to in the Conclusions (Section \ref{sec:conclusions}).

In the next subsection, we repeat similar experiments as those whose results are presented here, but with a different kind of circuit. 

\subsection{Model Performance: Quantum Machine Learning Circuits}

The previous sub-section evaluated the model's performance on quantum volume circuits. Next, we turn to the task of evaluating the model using a circuit used for quantum kernels, which evaluate a similarity measure $K(\mathbf{x},\mathbf{y})$ between two classical feature vectors $\mathbf{x}$, $\mathbf{y}$. Note in this section, synthetic values for $\mathbf{x}$ and $\mathbf{y}$ are used.

Given an encoding circuit $U(\mathbf{x})$, the corresponding quantum kernel circuit is $U(\mathbf{x})\circ U^{\dagger}(\mathbf{y})$. We focus on a particular encoding circuit on $n$ qubits, based on an encoding circuit introduced in \cite{havlivcek2019supervised}. The encoding circuit we use is given by
\begin{equation}
\label{eq:zzfm-circuit}
U(\mathbf{x}) =V(\mathbf{x})\circ H^{\otimes n},
\end{equation}
where $H^{\otimes n}$ is the Hadamard gate on all $n$ qubits, and
\begin{equation}
V(\mathbf{x}) = \text{Exp}\left(i\sum_{\mathbf{j} \in S}\left[ \phi_{\mathbf{j}}(\mathbf{x})\prod_{a \in \mathbf{j}}Z_{a} \right]\right).
\end{equation}
(Note that in \cite{havlivcek2019supervised}, the encoding circuit used is $ V(\mathbf{x}) \circ H^{\otimes n} \circ V(\mathbf{x})\circ H^{\otimes n}$.) Here the set $S$ indexes \emph{both} individual qubits, as well as pairs of them. 
The function $\phi_{\mathbf{j}}(\mathbf{x})$ is given by
\begin{equation}
    \phi_{\mathbf{j}}(\mathbf{x}) =
\begin{cases}
     \mathbf{x}_{j}~~~~~~~~~~~~~~~~~~~~~~~\text{single qubit $j$}\cr
     (\pi - \mathbf{x}_{j})(\pi - \mathbf{x}_{k})~~~~\text{qubit pair $j,k$}.
\end{cases}
\end{equation}

On the $j^{\mathrm{th}}$ individual qubit, $V(\mathbf{x})$ applies a phase rotation, with the phase being set by the value the $j^{\mathrm{th}}$ component of $\mathbf{x}$, $\mathbf{x}_{j}$. On a pair of qubits $j,k$, $V(\mathbf{x})$ applies an entangling $ZZ$ operation, with a phase set by $(\pi - \mathbf{x}_{j})(\pi - \mathbf{x}_{k})$.

Implicit in the notation above is the idea of an ``entangling strategy", which determines which pairs of qubits become entangled. In this work, we consider two strategies:
\begin{itemize}
    \item ``Linear", in which adjacent pairs of qubits are entangled: $S = \{0, 1, \cdots, n-1 \} \cup \{(0,1), (1,2), (2,3), \cdots, (n-2,n-1)\}$
    \item ``Full", in which all pairs of qubits are entangled:  $S = \{0, 1, \cdots, n-1 \} \cup \{(0,1), (0,2), (0,3), \cdots, (0,n-1), (1,2), \cdots, (n-2,n-1)\} $
\end{itemize}
In general, quantum kernel circuits are \emph{rectangular}: the total number of layers ($2D$) does not equal the circuit width ($n$). We use the \emph{aspect ratio} of the circuit, $a\equiv 2D/n$ to capture whether the kernel circuits are wide and shallow ($a < 1$), square ($a=1$), or narrow and deep  ($a > 1$).

Table \ref{tab:clops-exp-design-zzfm} shows the average performance of the model for kernel circuits with an aspect ratio $a=1$, and where $M=S=100$. (Note that here, the data is aggregated over circuits whose width varies between 2 and 6.) Similar behavior as Table \ref{tab:clops-exp-design-qv} is observed; namely, the model generally under-predicts job runtime. The degree to which the model does so depends on the entanglement structure of the encoding circuit. In particular, circuits with a ``linear" entanglement structure have a runtime ratio closer to 0 than those whose entanglement structure is ``full". From Figure \ref{fig:vol-area-depths}, we see the former family of circuits has a lower depth compared to quantum volume circuits of a similar volumetric area. This suggests the depth-dependent factor in the definition of $D_{\mathrm{eff}}$ in Equation \eqref{eq:eff-d} plays a significant role in the model's performance.

\begin{figure*}
\centering
\includegraphics[width=.9\textwidth]{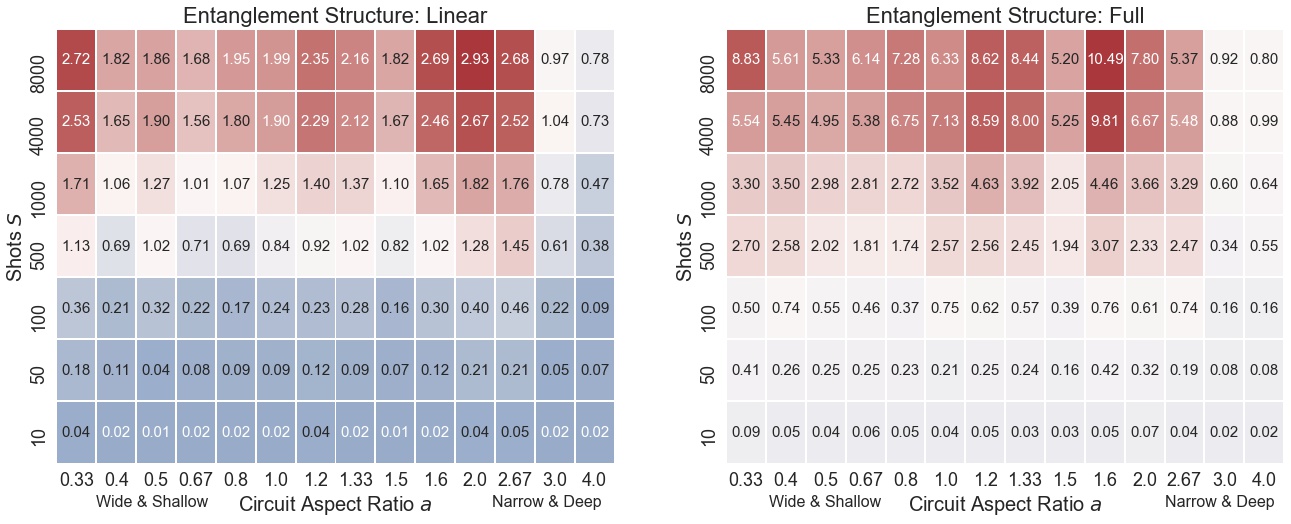}
\caption{\textbf{Effect of Changing the Number of Shots and Circuit Aspect Ratio (quantum kernel circuits).} The results of Table \ref{tab:clops-exp-design-zzfm} for the model's \emph{runtime ratio} are extended to \emph{rectangular} quantum kernel circuits (i.e., $a \neq 1$) and differing numbers of shots $S$, with $M=100$. Darker red colors indicate higher values of the runtime ratio, and darker blue colors, lower values.}
\label{fig:zzfm-estruc-loss}
\end{figure*}

\begin{table}
\centering

\caption{Model Performance for Square quantum kernel circuits (CLOPS-Like Job).}
\label{tab:clops-exp-design-zzfm}

\begin{tabular}{|l|c|c|c|c|}
\hline
{} & \multicolumn{2}{|l|}{\textbf{Loss} $L$} & \multicolumn{2}{l}{\textbf{Runtime Ratio} $r$} \\ \hline
\textit{Entanglement Structure} &  Full & Linear &          Full & Linear \\
\textbf{Backend}        &       &        &               &        \\ \hline
ibm\_hanoi      &  1.98 &  26.50 &          0.50 &   0.07 \\ \hline
ibmq\_guadalupe &  0.55 &   1.59 &          0.84 &   0.41 \\ \hline
ibmq\_toronto   &  0.13 &   4.84 &          1.01 &   0.25 \\ \hline
\end{tabular}

\end{table}

Figure \ref{fig:zzfm-estruc-loss} extends Table \ref{tab:clops-exp-design-zzfm} to include rectangular circuits, and to vary the number of shots. The behavior of the model is very similar as to what was seen for quantum volume circuits: namely, the model's runtime ratio decreases dramatically as $S\rightarrow 0$, and once $S$ is on the order of 500 or so, the ratio stabilizes. This behavior consistently occurs across a variety of circuit aspect ratios $a$, and is also consistent when the entanglement structure of the circuit is changed.

With respect to the circuit's aspect ratio, the model does better when the circuits are narrow and deep ($a > 1$) than wide and shallow ($a < 1$). This effect appears to be more pronounced for the ``full" entanglement structure, especially when $S$ is large. One way to understand this is that with the ``full" entanglement structure, every qubit is entangled with every other; for such circuits which are also wide, the depth of the circuit when transpiled to hardware could be quite large; this impacts the model's predictions via $D_{\mathrm{eff}}$.

\begin{figure*}[h!]
\centering
\includegraphics[width=.9\linewidth]{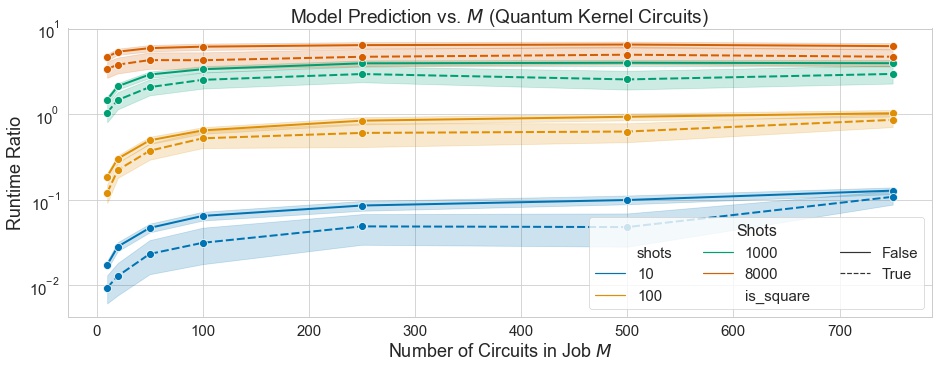}
\caption{\textbf{Assessing Accuracy of the Model as Measured by Runtime Ratio (quantum kernel circuits).} Incorporating all of the data collected for running quantum kernel circuits shows the model's performance is fairly stable with respect to $M$, and depends strongly on the number of shots $S$. Solid lines correspond to rectangular circuits ($a \neq 1$); dashed lines, square circuits ($a = 1$). Note: Lines correspond to the mean, and shading indicates a 95\% confidence interval, computed using the built-in APIs of the seaborn data visualization package \cite{Waskom2021}.}
\label{fig:zzfm-runtime-ratio}
\end{figure*}

Finally, we consider the impact of changing the number of circuits in the job, $M$. Figure \ref{fig:zzfm-runtime-ratio} shows the mean runtime ratio $r$ as a function of $M$, where the data is segregated on the number of shots, and whether $a=1$. The model's behavior is consistent for both square ($a=1$) and rectangular ($a\neq 1$) kernel circuits, and the mean runtime ratio is fairly stable across a wide range of values for $M$.

Taken together, Figures \ref{fig:zzfm-estruc-loss} and \ref{fig:zzfm-runtime-ratio} suggest that of the four parameters in the model, it is the number of shots $S$ and the number of effective quantum volume layers $D_{\mathrm{eff}}$ which play the most (and second-most) substantial role in influencing the model's performance, respectively. Given the assumptions of the model, this makes sense. $S$ enters multiplicatively in the model; as it goes down, the impact of fixed, shot-independent overheads becomes more important, but isn't explicitly captured by the model\footnote{As noted in Section \ref{sec:runtime-model}, this is an intentional choice, to avoid creating an unwieldy and over-parameterized model.}. The job's runtime is also impacted by how deep the circuits in the job are. The depth of the circuits is impacted both by the number of repetitions of the template and the entangling strategy, both of which impact $D_{\mathrm{eff}}$.

Having evaluating the model's performance on two kinds of circuits using synthetic data, we now turn to using the model to estimate runtimes for large data sets in a real-world context.

\section{Implications for Runtime on Practical Data Set Sizes}
\label{sec:model-extrapolation}

The prior section studied the model's performance. In this section, we use the model to examine the implications of running jobs for calculating quantum kernels where the underlying data set is both large and practical. The choice of the data set was influenced by the fact this work started  as part of a summer internship program offered by IBM and its Operations Risk Insights (ORI) organization\footnote{Operations Risk Insights (ORI) is an automated, comprehensive, and Watson-powered alert service which assesses employee safety, operations and natural disaster risk events to identify those posing the greatest threat of impact to the business continuity.}. Over the summer of 2022, ORI began incorporating into its capabilities a \emph{purely classical} model to predict flash floods. In parallel, the authors (and others, noted in the Acknowledgements) began exploring the use of a quantum-enhanced model through the use of quantum kernels \cite{QiskitqORIMedium}.

Flash floods are are a significant contributor to annual, weather-inflicted monetary losses.  They can be catastrophic to communities, infrastructure, and of course, people. Flash flood events are often unpredictable, making it hard to prepare for or mitigate their potential effects. For example, California's flooding rains and heavy snows which killed at least 17 people likely caused more than \$30 billion in damages and economic losses in January of 2023 \cite{CA-Flash-Flood}. Improved early warnings of flash floods thus can save lives and reduce economic losses. 

\begin{figure*}
\centering
\includegraphics[width=.9\textwidth]{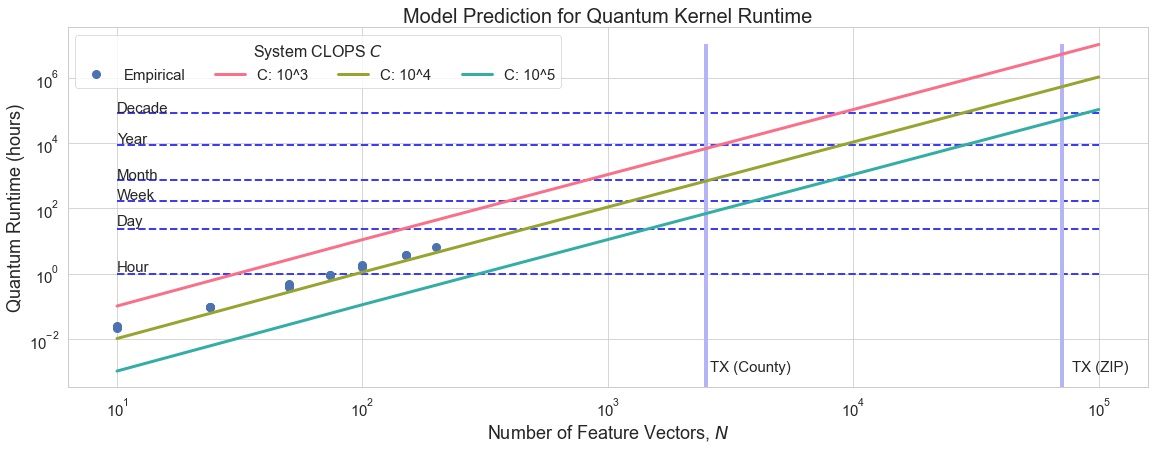}
\caption{\textbf{Extrapolating the Model to Large Data Sets.} Using the model to make predictions on runtime as a function of data set size $N$, with $S=4000$ and $D_{\mathrm{eff}}=2$. (Recall $M \sim \mathcal{O}(N^{2})$.) Colors indicate different system CLOPS values, $C$. The markers indicate empirical data collected using the Texas flash flood data set using the \emph{ibmq\_auckland} system, which -- as of the time of writing -- had a CLOPS of 2400. Details of those jobs are available in Appendix \ref{app:tx-data}. Vertical lines indicate the full data set size for Texas at the county and ZIP code level.}
\label{fig:model-extrapolation}
\end{figure*}

The ORI effort initially focused on flash flood prediction within the state of Texas, at two levels of geographic granularity: county level, and ZIP code level. At these two levels of granularity, the available data set had $N=2513$ records and $N=70571$ records, respectively. Although this number of records may be modest from a classical ML perspective, it is important to keep in mind that generating quantum kernels for both of these data sets requires running on the order of 3 million and 2.5 billion circuits, respectively. Utilizing the runtime model in Equation \eqref{eq:runtime-model}, we can roughly predict how long running those jobs would take.

Figure \ref{fig:model-extrapolation} plots the predictions of the model out to data set sizes encompassing both the Texas county and ZIP code data sets\footnote{For a given number of feature vectors $N$, the number of quantum kernel circuits $M = N(N-1)/2$.}. Here, specific values for both $D_{\mathrm{eff}}$ and $S$ are used; namely, $D_{\mathrm{eff}} = 2$ and $S=4000$. As we've seen in the previous section, the runtime will be impacted by both of these quantities. The primary focus of the figure is the impact of improving the speed of circuit execution (as measured by CLOPS, $C$).

Current system speeds are on the order of 1K. At such speeds, processing the Texas county data set would take on the order of approximately 1 year, and processing the Texas ZIP code data set would be infeasible for all practical purposes.

Recently, a demonstration of $C>10$K CLOPS has been made \cite{IBMSummitForbes}. At those system speeds, processing the Texas county data set could take on the order of months, and processing the Texas ZIP code data set would still remain infeasible.

Setting aside whether quantum advantage can be found for these particular data sets and the particular encoding circuit used, it is still useful to highlight how considerations from the overall flash-flood prediction workflow used by ORI would place constraints on the acceptable amount of runtime on quantum hardware, assuming  quantum-enhanced classifiers were deployed to the platform. That is, the ORI platform updates its flash flood predictions every 2 hours. If a quantum-enhanced classifier was incorporated into the platform, it would be necessary to refresh the kernel values within that time window. 
And while re-processing an entire data set may not be necessary, the implication from the model presented here is that, barring advances in the underlying algorithm itself, the runtime on quantum hardware would need to come down by several orders of magnitude in order for the quantum kernel part of the ORI platform to sustain the desired rate of updates for the model.

This highlights the quantum part of a quantum-enhanced workflow doesn't exist in isolation, and there are considerations which have nothing to do with quantum computing \emph{per se} which can impact the feasibility of deploying a quantum-enhanced approach to a classical workflow.

\section{Conclusions \& Discussion}
\label{sec:conclusions}

Quantum kernels are one particular quantum machine learning algorithm, in which classical ML models are enhanced by similarity measures computed by running quantum circuits on quantum systems. Given a data set of size $N$, $\mathcal{O}(N^{2})$ kernels need to be calculated. In this work, we studied the problem of modeling the runtime of a collection of circuits used to calculate quantum kernel values, and presented a predictive model to do so [Equation \eqref{eq:runtime-model}], based on a recently-introduced measure of the speed of quantum computers, CLOPS \cite{wack2021quality}.

We validated the model's performance by comparing its predictions against empirical runtime information, and found the model is most accurate when the job closely mimics those used to calculate the CLOPS of a given system. When the job being run is substantially different, the model's performance suffers. When the number of shots is small, the model consistently under-predicts runtime, due to the fact that, in reality, the software stack has fixed (and unavoidable!) overheads not accounted for by the model. When the number of shots is large, the model generally over-predicts runtime in a shot-dependent fashion. This suggests the model could be used -- to reasonable accuracy -- in a regime where the number of shots is modest, or large. Further, the model's performance is relatively stable as with respect to the number of circuits in the job, meaning it can be applied in the context of jobs with very large numbers of circuits.

We note here one of the main difficulties in making statement about the model as such is the degree to which the job execution path used to establish a system's CLOPS value differs from the one used here. This work leverages the Qiskit Runtime service for job execution, a service not currently used for CLOPS values. It would be interesting to re-consider the analysis presented here if it was, as we could then better understand whether the issues with the model's performance come from the model as such, or the particular execution path of the jobs.

By extrapolating the model to very large data set sizes (i.e., a number of feature vectors on the order of thousands and beyond), we find at current system speeds, processing such data sets would require a prohibitively large amount of runtime on quantum hardware. However, for smaller data set sizes, quantum kernels can be processed in a reasonable amount of time on today's systems. What's more, as noted in the Introduction, quantum advantage with quantum kernels has yet to be attained in a practical setting, meaning scaling up to larger data set sizes wouldn't be necessary right now for early users of quantum-enhanced models. That is, for small data set sizes, classical data scientists could already begin exploring quantum-enhanced, kernel based algorithms on real-world data, with circuit execution runtimes that enable interesting experimentation and work. In this sense, the speed of the hardware \emph{is not} an obstacle to data scientists and other early end-users of quantum-enhanced models to begin upskilling themselves today.

It is important to note this work \emph{does not} touch on the other practical or theoretical considerations necessary to substantiate a claim of quantum advantage. We make no claims -- nor dare speculate -- on whether improvements in job runtime would enable quantum advantage using the particular encoding circuit we studied, the particular quantum computing modality used (namely, superconducting qubits), and the particular data set considered.

The results of this work suggest 4 primary lines of additional research. First, there is a need to apply and validate the runtime model introduced here to a larger variety of circuits used for quantum machine learning. For example, ad-hoc (or ``hardware-efficient") circuits are used to encode data in a way with minimal circuit depth and for which their 2-qubit gates respect the connectivity of the qubits in the hardware. Studying a larger variety of circuits would provide more evidence of the regimes of validity of the model.

Second, hardware runtime could be further reduced through parallelization of the job across multiple QPUs. If the time on 1 QPU is $T$, parallelizing across $X > 1$ QPUs could reduce the total time to approximately $T/X$. As more quantum systems come online, the feasibility of doing this parallelization becomes higher\footnote{Note this approach ignores any latency effects, the overhead of the software orchestrating the parallelization, and the potentiality of the parallelized jobs being sent to different queues, each with their own queue behavior.}. Further, multiple quantum kernel circuits could be executed on the same chip, assuming a sufficient number of qubits is available. This would provide another level of parallelization.

Third, one of the most straightforward ways to decrease job execution is to reduce the number of shots $S$. Doing so comes with the cost of increasing the shot noise of the estimated kernel values. A close collaboration with classical ML scientists and practitioners looking at kernelized ML algorithms with robust performance guarantees in the face of noisy kernel values would be fruitful, and could help the quantum ML research community understand what the practical upper bounds on $S$ might be, both in the context of quantum-enhanced support vector machines, and other ML algorithms. For example, recent work has shown that in order for an SVM to have a generalization error at most $\epsilon$ when trained on a data set of size $N$, the total number of shots required per kernel entry scales as $S \sim \mathcal{O}(N^{8/3}/\epsilon^{2})$ \cite{gentinetta2022complexity}. In turn, this implies a runtime -- across the entire data set -- of $\mathcal{O}(N^{2})*\mathcal{O}(N^{8/3}/\epsilon^{2}) * D_{\mathrm{eff}}/C = \mathcal{O}(N^{4.67}D_{\mathrm{eff}}/(C\epsilon^{2}))$. This is a rather unfavorable scaling with respect to $N$ in practice, and motivates exploring regimes wherein small amounts of training data are required, and algorithms which can tolerate relatively large amounts of error in the estimated kernel entries.

Fourth, the notion of ``effective number of quantum volume layers of a circuit" should be studied in more depth. We presented one definition [Equation \eqref{eq:eff-d}]; others are possible. In particular, the definition of $D_{\mathrm{eff}}$ introduced here was particular to quantum kernel circuits; defining one which could be applied across a wider family of circuits would be useful.

In sum, this work showed it is possible to model job execution time using a holistic measure of the speed of quantum systems. This model has four parameters: number of circuits $M$, number of shots $S$, system CLOPS $C$, and number of effective quantum volume layers $D_{\mathrm{eff}}$. Although simple, we showed this model can be used -- with reasonable accuracy -- to predict job execution time, especially in a regime where the number of shots is large. We encourage end-users of quantum computing systems to leverage this model for analyzing the quantum-enhanced portion of their workflows, and for quantum computing applications researchers to find ways to apply it to other applications of quantum algorithms beyond quantum kernels.

\section{Acknowledgements}

We acknowledge prior collaborative contributions from the other IBM ORI Extreme Blue Interns for Summer 2022: Chelsea Zackey, Christopher Moppel, and Samantha Anthony. Further, we acknowledge the support of other ORI Exterme Blue mentors, including Bhanwar Gupta, Chester Karwatowski,  Rinku Kanwar, Mallikarjun Motagi and Ayush Kumar.  In addition, we acknowledge the support of the IBM Extreme Blue program, as well as Dr. Liliana Horne of IBM's Global Chief Data Office. TLS thanks Drs. Paul Nation, Omar Shehab, and Stefan W\"{o}rner  for feedback on earlier versions of this manuscript. JW thanks Fausto Palma of the IBM CIO Supply Chain and Technology Systems group for his gracious support. Finally, we acknowledge the use of IBM Quantum services for this work. The views expressed are those of the authors, and do not reflect the official policy or position of IBM or  IBM Quantum.

\appendix

\section{Real-World Results: Methods and Details}
\label{app:tx-data}
This appendix describes the methods and workflows used to generate the empirical results presented in Figure \ref{fig:model-extrapolation}. These workflows were built in a broader context of creating an end-to-end pipeline for training classical and quantum-enhanced models, leveraging state-of-the-art, cloud-based tools. In particular, the workflows were built using Kubeflow \cite{kubeflow} running on the IBM Cloud Kubernetes Service, to manage the complexity of both the classical and quantum machine learning experimental workflows. Kubeflow is an open source toolkit and a de-facto standard for building, experimenting with, and deploying ML pipelines to various environments for development, testing, and production-level model serving, on containerized environments such as Red Hat OpenShift \cite{openshift} and vanilla Kubernetes \cite{k8s}. Within Kubeflow are Kubeflow Pipelines (KFP), which is a “platform for building and deploying portable, scalable machine learning (ML) workflows based on Docker containers”. Each KFP step or component is containerized, with the ability to share and track results and associated experiment artifacts between components, while allowing independent, long-running steps to proceed in parallel.

\begin{figure}
\centering
\includegraphics[width=\linewidth]{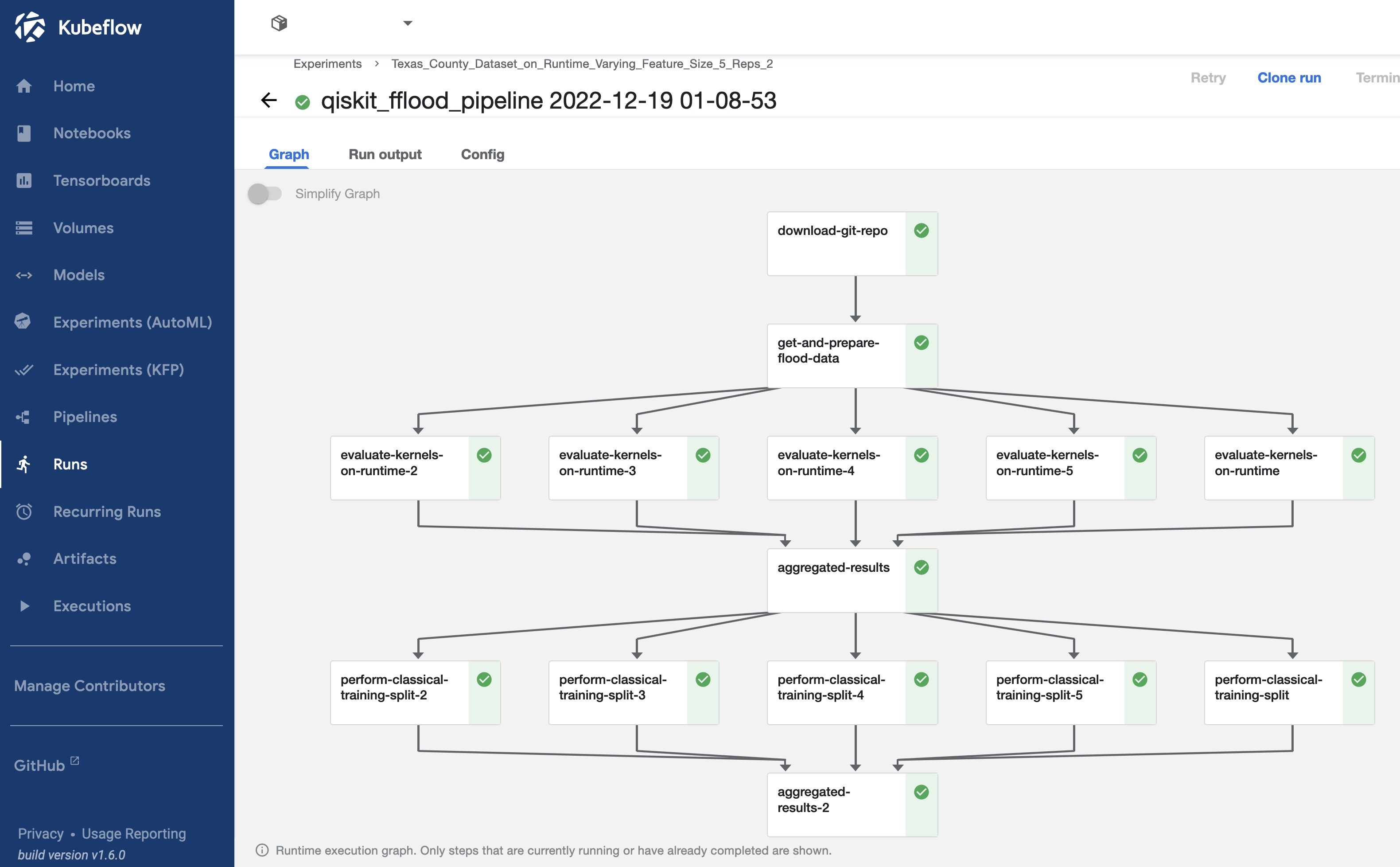}
\caption{\textbf{Kubeflow Dashboard showing a completed Kubeflow Pipeline Run.} Kubeflow allows for large, complex experiments to be run as independent and parallel steps. (Time proceeds from top to bottom.) The workflow shown above workflow proceeds as follows: source code download, data preparation, generating quantum kernel values, aggregating Qiskit Runtime job results, generating classical kernel values, and model training \& analysis.}
\label{fig:completed-kubeflow-pipeline}
\end{figure}

The end-to-end Kubeflow pipeline consisted of the following steps:
\begin{enumerate}
\item Initialization: Obtaining the latest source code binaries from Github
\item Data preparation: Performing feature selection and data resizing.  
\item Quantum kernel generation: create the jobs needed to calculate quantum kernel values, and send them to IBM Quantum systems.
\item Aggregate Qiskit Runtime job results: extract empirical runtime information and a quantum kernel matrix from the job results
\item Classical kernel generation: calculate a classical kernel (RBF kernel) for the data set generated in Step 2.
\item Model training and analysis: train 2 SVMs (one for each kernel matrix), and evaluate their accuracy.
\end{enumerate}

An example pipeline -- showing the launching of 5 independent quantum and classical kernel generation tasks -- is given in Figure \ref{fig:completed-kubeflow-pipeline}. A major benefit of using Kubeflow for running the experiment done in this work is the ability to parallelize the workflow across multiple splits, where each split can consist of independent data sets. In addition, pipeline runs are automated and asynchronous, on a managed cloud environment (vs., e.g., running manually on a standalone machine). As a result, a very large experiment can be split into multiple independent ones, meaning the failure of any one sub-experiment does not impact whether other sub-experiments fail. This also allows for an easy reboot/restart of the failed sub-experiments. In addition, because of the cloud-based nature of Kubeflow, long-running experiments (e.g., several hours) can be easily handled, due to the fact the orchestration of the work is handled via the cloud. Finally, the use of splits allows for more usage of Qiskit Runtime compute resources as they become available, by, e.g., sending different splits to different systems.

We now provide brief descriptions of some of the steps above.

For step 2, the real-world data sets used consisted of 38 features, and was constructed out of long-term flash flood records and historical analysis from the following sources: 
\begin{itemize}
\item National Oceanic and Atmospheric Administration (NOAA), for historical precipitation data
\item The Weather Channel (TWC), for hourly atmospheric and precipitation data
\item Multi-Resolution Land Characteristics Consortium (MRLC), for land surface data
\item  US Geological Survey (USGS), for regional land classification
\end{itemize}
The particular dataset used here is one generated for the state of Texas at the county level, which had  $N=2513$ records. For the data preprocessing, classical principal component analysis (PCA) was used to perform feature reduction to go from the initial 38 features to the statistically most significant 2, 3, 5, and 7 features. This allowed for a study of the impact on model accuracy as the number of features was changed. For the data points in Figure \ref{fig:model-extrapolation}, the two most significant features -- \textit{PrecipAmountAvg} and \textit{RelativeHumidityAvg} -- were used.

Because flash floods represented only 3\% of the data set, caution was needed during data preparation to avoid issues that are typical of highly imbalanced datasets. When attempting resize the dataset from the initial $N=2513$ records to smaller batches of $N=10, 25, 50, 75, 100, 150, 200$ the Imbalanced Learn RandomUnderSampler  \cite{RandomUnderSampler} was used ensure we maintained an appropriate representation of flash floods in the training dataset. Note that both the feature reduction and the data resizing are done each time our experimental pipelines are run, as they are computationally easy.

For step 3, the code used to generate jobs consisting of quantum kernel circuits was based on the open source, quantum kernel library in Qiskit Machine Learning project \cite{QiskitQMLGithub}, the \textsc{compute\_overlap}, \textsc{compute\_circuit}, and \textsc{evaluate} methods in particular. These functions were modified to include calls to the Qiskit Runtime APIs to facilitate the extraction of job execution information, as well as the quantum kernel matrix itself (step 4). Jobs were run on the \emph{ibmq\_auckland} system, using a dedicated reservation mode made available via the IBM Quantum Platform. The \emph{ibmq\_auckland} system is a 27 qubit machine, with a quantum volume of 64, and CLOPS of 2400. 

For step 5, the choice of RBF kernel was motivated by prior work from the authors and other collaborators \cite{QiskitqORIMedium}, which showed the RBF kernel yielded the best balanced accuracy and F1 score compared to other classical kernel functions and model approaches for the flash flood data set. This step is part of the pipeline since it is not computationally intensive, and provided a classical benchmark against which to compare a quantum-enhanced classifier.

\newpage 

\printbibliography
\end{document}